\documentclass[fleqn]{article}

\usepackage{amsmath,amssymb}
\usepackage{epsfig}
\usepackage{graphicx}
\usepackage[top=0.75in, bottom=0.75in, left=0.75in, right=0.75in, dvips]{geometry}

\begin{document}

\title{{\sf Exact One Loop Running Couplings in the Standard Model}}
\author{
F. A.\ Chishtie\thanks{Department of Applied
Mathematics, The University of Western Ontario, London, ON, N6A
5B7, Canada}, M.D.\ Lepage\thanks{St.\ Peter's College, Muenster, SK, Canada}, D.G.C.\ McKeon\thanks{Department of Applied
Mathematics, The University of Western Ontario, London, ON, N6A
5B7, Canada}, T.G.\ Steele\thanks{Department of Physics and
Engineering Physics, University of Saskatchewan, Saskatoon, SK,
S7N 5E2, Canada}, I. Zakout\thanks{Department of Applied
Mathematics, The University of Western Ontario, London, ON, N6A
5B7, Canada}}

\maketitle

\begin{abstract}
Taking the dominant couplings in the standard model to be the
quartic scalar coupling, the Yukawa coupling of the top quark, and
the SU(3) gauge coupling, we consider their associated running
couplings to one loop order. Despite the non-linear nature of the
differential equations governing these functions, we show that
they can be solved exactly. The nature of these solutions is
discussed and their singularity structure is examined. It is shown that for a sufficiently small Higgs mass,
the quartic scalar coupling decreases with increasing energy scale
and becomes negative, indicative of vacuum instability. This behavior changes for a Higgs mass
greater than 168 GeV, beyond which this couplant increases with
increasing energy scales and becomes singular prior to the ultraviolet (UV) pole of the Yukawa coupling.
Upper and lower bounds on the Higgs mass corresponding to new physics at the TeV scale are obtained and compare favourably with 
the numerical results of the one-loop and two-loop analyses with inclusion of electroweak couplings. 
\end{abstract}

It is well understood that removing infinities arising in quantum
field theory when evaluating perturbative corrections to classical
interactions induces a dependence of the physical couplings on the
energy scale $\mu$ of the process being considered. The most
surprising consequence of this dependence is asymptotic
freedom, the decrease in the non-Abelian gauge coupling with
increasing energy scale. Clearly, the asymptotic behaviour of the
running couplings is of great theoretical and
phenomenological interest.

The evolution of these running couplings is dictated by a set of
non-linear ordinary differential equations in which the various
couplings are inextricably linked. This can change a naive
expectation of their behaviour based on ignoring this linkage. For
example, the quartic scalar self-coupling $\lambda$ of a pure
$O(N)$ scalar theory diverges with increasing energy scale, but
this need not be true if the coupling is affected by the
interaction of this scalar with other fields. In this paper, we 
 demonstrate via an exact solution that this quartic scalar
coupling $\lambda$ in the Standard Model has an asymptotic
behaviour and singularity structure that is strongly affected by the top-quark coupling
$g_t$ and the SU(3) gauge coupling $g_3$, the dominant couplings
of the Standard Model at the weak scale.  We show this by
explicitly solving in closed form  the one-loop equations that
govern how $\lambda$, $g_t$ and $g_3$ evolve with $\mu$. The
importance of having an explicitly analytic solution to a set of
non-linear coupling equations has been emphasized in
\cite{Nicolai}. In fact, the asymptotic behaviour and singularity structure of $\lambda$
becomes contingent upon the boundary conditions to these
equations, and those boundary conditions in turn depend on the
Higgs mass $M_H$. For $M_H < 168$ GeV we find that $\lambda$
decreases with increasing $\mu$ and eventually goes negative; for
$M_H > 168$ GeV,  $\lambda$ will be shown to remain positive up to its singularity. This
provides an independent way of using the renormalization group
equation to analyze the sensitivity of the Standard Model on $M_H$
that is complementary to  numerical approaches 
\cite{2}. Furthermore, having an exact solution for these
running couplings will provide a way of determining the effective
potential at leading-log order in the conformal limit of the
Standard Model using the method of characteristics \cite{1}, which
in turn has cosmological implications \cite{Alex}.

The one-loop equations that determine how $\lambda$, $g_t$ and
$g_3$ depend on $\mu$ are \cite{Sher}:
\begin{equation}
\dot{x}=\frac{9}{4}x^2-4xz
\label{rg_eq1}
\end{equation}
\begin{equation}
\dot{y}=6y^2+3xy-\frac{3}{2}x^2
\label{rg_eq2}
\end{equation}
\begin{equation}
\dot{z}=-\frac{7}{2}z^2
\label{rg_eq3}
\end{equation}
where $x=\frac{g_{t}^2}{4\pi^2}$, $y=\frac{\lambda}{4\pi^2}$,
$z=\frac{g_{3}^2}{4\pi^2}$ and 
the dot denotes the derivative with respect to
$t=\log(\mu)$. Here $\mu$ is the renormalization-induced mass parameter in the
theory.\footnote{It does not appear to be possible to obtain
an exact solution by applying computer algebra programs
directly to the differential equations for these one loop
couplings.}

Eq.~(\ref{rg_eq3}) can be solved immediately to give
\begin{equation}
z(t)=\frac{z_0}{1+\frac{7}{2}z_0(t-t_0)} \label{z_sol}
\end{equation}
where $z_0=z(t_0)$ is a boundary value for $z(t)$. Together,
Eq.~(\ref{rg_eq1}) and Eq.~(\ref{rg_eq3}) give
\begin{equation}
\frac{dx}{dz}=-\frac{2}{7}\left[\frac{9}{4}\left(\frac{x}{z}\right)^2-4\left(\frac{x}{z}\right)\right]
\end{equation}
which, if $x=zw$, becomes
\begin{equation}
z\frac{dw}{dz}=-\frac{9}{14}w^2+\frac{1}{7}w
\label{x_trans}
\end{equation}
Eq.~(\ref{x_trans}) can be immediately integrated to give
\begin{equation}
\left(\frac{z}{z_0}\right)^{\frac{1}{7}}=\left(\frac{w}{w-\frac{2}{9}}\right)\left(\frac{w_0-\frac{2}{9}}{w_0}\right)
\label{x_sol}
\end{equation}
where $w_0=w(t_0)=\frac{x(t_0)}{z(t_0)}$.
If we define
$K={z_0}^{\frac{1}{7}}\left(\frac{w_0-\frac{2}{9}}{w_0}\right)$,
then Eq.~(7) leads to
\begin{equation}
x(t)=\frac{2z(t)}{9\left[1 - K z(t)^{-{\frac{1}{7}}}\right]}
\label{x_sol2}
\end{equation}

We now must examine how the scalar couplant, $y$, evolves with
$t$. Dividing Eq.~(2) by Eq.~(3), we obtain
\begin{equation}
\frac{dy}{dz}=-\frac{2}{7}\left[6\left(\frac{y}{z}\right)^2+3\left(\frac{x}{z}\right)\left(\frac{y}{z}\right)-\frac{3}{2}\left(\frac{x}{z}\right)^2\right]
\label{y_trans1}
\end{equation}
From  Eq.~(7), we can put
\begin{equation}
w=\frac{x}{z}=\frac{p}{\tau+q}
\label{w_def}
\end{equation}
where $\tau=z^{-\frac{1}{7}}$, $p=-\frac{2}{9K}$ and
$q=-\frac{1}{K}$, so that if $u=\frac{y}{z}$, then Eq.~(9) becomes
\begin{equation}
\tau\frac{du}{d\tau}-12u^2-(6w+7)u=-3w^2 \label{y_trans2}
\end{equation}
Eq.~(\ref{y_trans2}) is a Ricatti equation; the transformation
\begin{equation}
u=-\frac{\tau\frac{dv}{d\tau}}{12v}
\label{ricatti1}
\end{equation}
brings it to the form
\begin{equation}
\frac{d^2v}{d\lambda^2}+\left(\frac{r}{\lambda+1}-\frac{r+6}{\lambda}\right)\frac{dv}{d\lambda}
+\frac{r^2}{\lambda(\lambda+1)}\left(\frac{1}{\lambda+1}-\frac{1}{\lambda}\right)v=0
\label{ricatti2}
\end{equation}
where $\tau=\lambda q$ and $r=\frac{6p}{q}=\frac{4}{3}$. A
rescaling of $v$ so that
\begin{equation}
v=\lambda^{\alpha}(\lambda+1)^{\beta}f
\label{ricatti3}
\end{equation}
leads to
\begin{equation}
\frac{d^2f}{d\lambda^2}+\left(\frac{2\alpha-\frac{22}{3}}{\lambda}+\frac{2\beta+\frac{4}{3}}{\lambda+1}\right)\frac{df}{d\lambda}
+\left(2\alpha\beta+\frac{4}{3}\alpha-\frac{22}{3}\beta+\frac{32}{9}\right)\left(\frac{1}{\lambda}-\frac{1}{\lambda+1}\right)f=0
\label{ricatti4}
\end{equation}
provided the conditions
\begin{eqnarray}
\alpha^2-\frac{25}{3}\alpha-\frac{16}{9}=0 \\\beta^2
+\frac{1}{3}\beta -\frac{16}{9}=0
\end{eqnarray}
are used to eliminate terms proportional to $\frac{f}{\lambda^2}$
and $\frac{f}{(1+\lambda)^2}$ which arise from Eq.~(13). Taking
the solutions to Eqs.~(16) and (17) to be
\begin{eqnarray}
\alpha=\frac{25+\sqrt{689}}{6} \\
\beta= \frac{-1-\sqrt{65}}{6}
\end{eqnarray}
Eq.~(\ref{ricatti4}) is of the form of a hypergeometric
differential equation whose independent solutions about
$\lambda=0$ when $|-\lambda|< 1$, are
\begin{equation}
f=F(a,b;c;-\lambda)
\end{equation}
and
\begin{equation}
f=(-\lambda)^{1-c}F(a-c+1,b-c+1;2-c;-\lambda)
\end{equation}
provided
\begin{equation}
ab=2\alpha\beta+\frac{4\alpha}{3}-\frac{22\beta}{3}+\frac{32}{9}
  =\frac{1}{18}\left[161-3\sqrt{65}+3\sqrt{689}-13\sqrt{265}\right]
\label{ab1}
\end{equation}

\begin{equation}
a+b=2\alpha+2\beta-7
  =1+\frac{1}{3}\left(\sqrt{689}-\sqrt{65}\right)
\label{ab2}
\end{equation}
\begin{equation}
c=2\alpha-\frac{22}{3}
  =1+\frac{1}{3}\sqrt{689}
\end{equation}
Solving Eqs.~(\ref{ab1}) and (\ref{ab2}) we obtain
$a=4+\frac{1}{6}(\sqrt{689}-\sqrt{65})$ and
$b=-3+\frac{1}{6}(\sqrt{689}-\sqrt{65})$.

The choice of signs in the roots appearing in Eqs. (18) and (19)
ensures that $c-a-b=-\frac{1}{3}-2\beta$ is positive so that
$F(a,b;c;z)$ is well defined as $z\rightarrow 1^{-}$ as
$\lim_{z\rightarrow
1^{-}}F(a,b;c;z)=\frac{\Gamma(c)\Gamma(c-a-b)}{\Gamma(c-a)\Gamma(c-b)}$.
As $t$ increases, so that
$-\frac{\tau}{q}=\left[1+\frac{7}{2}(t-t_0)\right]^{\frac{1}{7}}\left(1-\frac{2z_0}{9x_0}\right)$
exceeds one in magnitude, we must analytically continue the
solutions of Eqs. (20, 21) using the standard results for
analytically continuing the hypergeometric function $F(a,b;c;z)$
into a domain in which $|z|\geq 1$. For Re $z < \frac{1}{2}$, one
can use the result

\begin{equation}
F(a,b;c;z)=(1-z)^{a}F\left(a,c-b;c;\frac{z}{1-z}\right)\nonumber
\end{equation}
while for $|1-z|<1$ we have,
\begin{eqnarray}
F(a,b;c;z)=\frac{\Gamma(c)\Gamma(c-a-b)}{\Gamma(c-a)\Gamma(c-b)}F\left(a,b;1+a+b-c;1-z\right)+\nonumber\\
(1-z)^{c-a-b}\frac{\Gamma(c)\Gamma(a+b-c)}{\Gamma(a)\Gamma(b)}F\left(c-a,c-b;1-a-b+c;1-z\right)\nonumber
\end{eqnarray}
These extensions are valid for the values of $a$,$b$ and $c$ given
by Eqs. (22-24). However, for the values of $-\frac{\tau}{q}$ we
consider, Eqs. (20-21) are valid solutions to Eq. (15). We also
note that according to Eq. (8), 
\begin{equation}
x(t)=\frac{2z(t)}{9\left[1+\frac{\tau}{q}\right]}
\nonumber
\end{equation}
so that $x(t)$ develops a pole when $-\frac{\tau}{q} \rightarrow
1$.

Together Eqs.~(12),(14) and (20) show that
\begin{equation}
y=-z\frac{\tau\frac{d}{d\tau}\left[\tau^{\alpha}(\tau+q)^{\beta}\left(F(a,b;c;-\frac{\tau}{q})+C\left(-\frac{\tau}{q}\right)^{1-c}F(a-c+1,b-c+1;2-c;-\frac{\tau}{q})\right)\right]}
{12\tau^{\alpha}(\tau+q)^{\beta}\left(F(a,b;c;-\frac{\tau}{q})+C\left(-\frac{\tau}{q}\right)^{1-c}F(a-c+1,b-c+1;2-c;-\frac{\tau}{q})\right)}
\end{equation}
where, as noted above, $\tau=z^{-{\frac{1}{7}}}$, $q$ and $C$ are
constants of integration (with $q=-\frac{1}{K}$) and
$a$,$b$,$c$,$\alpha$,$\beta$ are defined by Eqs.~(18,19,22--24). The
derivatives in Eq.~(25) can be computed using the equation
\begin{equation}
\frac{d}{dz}F(a,b;c;z)=\frac{ab}{c}F(a+1,b+1;c+1;z)~,
\end{equation}
resulting in the expression
\begin{equation}
\begin{split}
y=& -\frac{z\alpha}{12}-\frac{z\beta}{12}\frac{\frac{\tau}{q}}{\left(1+\frac{\tau}{q}\right)}
\\&
+\frac{z}{12}\frac{\tau}{q}
\frac{\left[\frac{ab}{c}F(a+1,b+1;c+1;-\frac{\tau}{q})+C(1-c)(-\tau/q)^{-c}F(a-c+1,b-c+1;2-c;-\frac{\tau}{q})\right]}{\left[F(a,b;c;-\frac{\tau}{q})+C\left(-\frac{\tau}{q}\right)^{1-c}F(a-c+1,b-c+1;2-c;-\frac{\tau}{q})\right]}
\\
&+\frac{z}{12}\frac{\tau}{q}
\frac{\left[\frac{(a-c+1)(b-c+1)}{2-c} C (-\tau/q)^{1-c}F(a-c+2,b-c+2;3c;-\frac{\tau}{q})\right]}{\left[F(a,b;c;-\frac{\tau}{q})+C\left(-\frac{\tau}{q}\right)^{1-c}F(a-c+1,b-c+1;2-c;-\frac{\tau}{q})\right]}~.
\end{split}
\end{equation}
 indicating that the Yukawa and scalar couplings share an UV pole at $-\tau/q=1$.  In addition, the scalar couplant can also develop poles
when the denominator shared by the last two terms is zero. Hence, the analytic
solution in the scalar couplant [Eq.\ (27)] additionally 
points to a non-trivial pole structure which builds upon other
exact solutions in in the literature \cite{Lindner,2},
thereby increasing our understanding of the behaviour of running
couplings in the Standard Model, most specifically in the case of
a scalar field in the presence of both a gauge and Yukawa
coupling.

With initial values of $y_0$ being taken to be fixed by tree level
relation $y(\phi_0)=\frac{M_H^2}{8 \pi^2 \phi_0^2}$,
$x_0=x(\phi_0)=0.0253$, $z_0=z(\phi_0)=0.0329$ \cite{1} (where
$M_H$ denotes the mass of a Standard Model Higgs boson and
$\phi_0=246.2$ GeV denotes the vacuum expectation value of the
scalar field $\phi$), the integration constant $C$ is obtained
from Eq.~(25)  in terms of $M_H^2$:
\begin{equation}
C={\frac { 4.862287086+ 0.00003014398478{{M_H}}^{2}}{ 1.236892814
- 0.00004219469394{{ M_H}}^{2}}}
\end{equation}
where all quantities are in GeV units.

With these initial values, the energy-dependence of the solution (27) for the scalar
couplant $y$ can be examined.\footnote{We have verified that our analytic results are in agreement with a numerical solution of Eqs.~(1)--(3).} For sufficiently small Higgs masses, 
the couplant decreases with increasing energy and
eventually turns negative, indicating a vacuum instability in the
theory.  When the Higgs mass
becomes sufficiently large this
behavior is seen to dramatically change from being a decreasing
function to an increasing function of $\mu$,  ultimately approaching a UV
pole.   The emergence of a pole in the scalar coupling prior to the pole in the Yukawa coupling indicates that the 
denominators in the last two terms of (28) become zero prior to $-\tau/q=1$.
This indicates limits on the applicability of perturbation theory
for a sufficiently heavy Higgs mass. This is also consistent with
the behaviour of the quartic scalar couplant reported in
\cite{Lindner,2,3}.

The value of $M_H$ representing the boundary between these two scenarios corresponds to the case where the pole of the Yukawa coupling [and the second term in (27)] coincides with the shared pole of the last two terms in (27). In other words, this bound on $M_H$ results in a scalar coupling 
that remains positive until it 
becomes singular at the same energy scale as the Yukawa coupling.  Setting the denominator of the last two terms in (27) to zero at the 
point $-\tau/q=1$  leads to the following expression for $C$
\begin{equation}
C=-\frac{\Gamma(c)\Gamma(1-a)\Gamma(1-b)}{\Gamma(c-a)\Gamma(c-b)\Gamma(2-c)}
\end{equation}
corresponding via (28) to $M_H=168.67\,{\rm GeV}$.\footnote{The value $M_H=168.67\,{\rm GeV}$ obtained from the solution of Eqs.~(28) and (29) has also been verified by numerical exploration of Eq.~(25).} 
Although the electroweak couplings have been ignored in the RG equations (1)--(3), this approximation does not have a significant impact on the numerical value of the boundary value of $M_H$. If one augments the one-loop RG equations to include the electroweak couplings, their numerical solution would lead to a value of $M_H=160.8\,{\rm GeV}$ above which $\lambda$ remains positive up to its singular value.

Remarkably, the phenomenological implications of the one-loop analytic expression for the scalar coupling are quite consistent with more detailed one-loop and 
two-loop numerical results including the electroweak gauge couplings.  For example, if we assume a $1\,{\rm TeV}$ scale at which the Standard Model breaks down, then the naive vacuum stability requirement $y(1\,{\rm TeV})>0$ leads via (27) to the bound $M_H>75\,{\rm GeV}$.  By comparison,  augmenting the one-loop RG equations with electroweak couplings results in $M_H>72\,{\rm GeV}$ from the numerical solution.   Similarly, an upper bound on $M_H$ corresponding to a pole at $1\,{\rm TeV}$ in the last two terms of (27) results in $M_H<740\,{\rm GeV}$, 
while the one-loop numerical solution augmented by electroweak couplings results in $M_H<744\,{\rm GeV}$.
The upper and lower bounds on $M_H$ resulting from our analytic solution are also in good agreement with the  full two-loop numerical analyses \cite{2}.  In particular, our bounds on $M_H$ would not appear out of place in Figure 2 of \cite{plot1} or Figure 3 of \cite{plot2}.  

The exact solution of the RG equations in the one-loop approximation with the dominant   electroweak scale couplings ($g_t$, $\lambda$, and $g_3$) thus appears to reproduce the essential features of Higgs mass-bound phenomenology.  We speculate that the underlying singularity structure manifested by the analytic solution (27) is responsible for this concordance. 

Finally, we note that knowing the exact behaviour of these one loop
couplants would in principle lead to an exact expression for the one
loop effective potential in the Standard Model when it is computed
using the method of characteristics \cite{1}.

\section*{Acknowledgements}
NSERC provided financial support. Roger Macleod had helpful
advice.


\begin{thebibliography}{99}


\bibitem{1}V.\ Elias, R.B.\ Mann, D.G.C.\ McKeon and T.G.\ Steele, Nucl.\ Phys.\ B 678 (2004) 147; (Erratum) Nucl.\ Phys.\ B 703 (2004) 413.

\bibitem{Nicolai}  K.~A.~Meissner and H.~Nicolai,``Conformal symmetry and the standard model,'' [arXiv:hep-th/0612165].

\bibitem{Lindner} M.\ Lindner, Z.~Phys.~C31 (1986) 295; \\
B.\ Grzadkowski and M.\ Lindner, Phys.~Lett.~B193 (1987) 71;\\
B.\ Grzadkowski, M.\ Lindner, S.\ Theisen, Phys.~Lett.~B 198 (1987) 64.

\bibitem{2}
J.~Kubo, K.~Sibold and W.~Zimmermann, Nucl.~Phys.~B259 (1985) 331;
Jonathan Bagger, Savas Dimopoulos and  Eduard Mass\'o, Phys.~Rev.~Lett.~55 (1985)  
T.~Hambye and K.~Riesselmann, Phys.\ Rev.\ D {\bf 55} (1997) 7255.

\bibitem{Alex} A.~Buchel, F.~A.~Chishtie, V.~Elias, K.~Freese, R.~B.~Mann, D.~G.~C.~McKeon and T.~G.~Steele, JCAP {0503} (2005) 003.

\bibitem{Sher} M.~Sher, Phys.\ Rept.\  {\bf 179}, 273 (1989).

\bibitem{3} Joseph D.\ Lykken,  ``The standard model: Alchemy and astrology", [arXiv:hep-ph/0609274].

\bibitem{bound} 
G.~ Altarelli and G.~Isidori, Phys.~Lett.~B337 (1994) 141;\\
J.A.~Casas, J.R.~Espinosa and M.~Quiros, Phys.~Lett.~B382 (1996) 374.
 
\bibitem{plot1} Marcela Carena and Howard E. Haber
Prog.~Part.~Nucl.~Phys.~ 50 (2003) 63.

\bibitem{plot2}  
Kurt Riesselmann, ``Limitations of a Standard Model Higgs Boson'', [arXiv:hep-ph/9711456]




\end{thebibliography}
\end{document}